\documentclass[acp,pre]{copernicus}
\usepackage{color}
\usepackage{soul}

\begin{document}
\nolinenumbers
\title{Estimating scalar turbulent fluxes with slow-response sensors in the stable atmospheric boundary layer}


\Author[1,2]{Mohammad}{Allouche}
\Author[1,3]{Vladislav I.}{Sevostianov}
\Author[1]{Einara}{Zahn}
\Author[1,3]{Mark A.}{Zondlo}
\Author[5]{Nelson Luís}{Dias}
\Author[4]{Gabriel G.}{Katul}
\Author[6]{Jose D.}{Fuentes}
\Author[1]{Elie}{Bou-Zeid}

\affil[1]{Department of Civil and Environmental Engineering, Princeton University, Princeton, New Jersey, USA}
\affil[2]{Lawrence Livermore National Laboratory, Livermore, California, USA}
\affil[3]{Princeton Materials Institute, Princeton University, Princeton, New Jersey, USA}
\affil[4]{Department of Civil and Environmental Engineering, Duke University, Durham, North Carolina, USA}
\affil[5]{Department of Environmental Engineering, Federal University of Paraná, Curitiba, PR, Brazil}
\affil[6]{Department of Meteorology and Atmospheric Sciences, The Pennsylvania State University, University Park, Pennsylvania, USA}




\correspondence{Elie Bou-Zeid (ebouzeid@princeton.edu)}

\runningtitle{REA}

\runningauthor{Allouche et al.}

\received{}
\pubdiscuss{} 
\revised{}
\accepted{}
\published{}


\firstpage{1}

\maketitle

\begin{abstract}
Conventional and recently developed approaches for estimating turbulent scalar fluxes under stable conditions are evaluated.  The focus is on methods that do not require fast scalar sensors such as the relaxed eddy accumulation (REA) approach, the disjunct eddy-covariance (DEC) approach, and a novel mixing length parametrization labelled as A22. Using high-frequency measurements collected from two contrasting sites (Utqiagvik, Alaska and Wendell, Idaho "during winter"), it is shown that the REA and A22 models outperform the conventional Monin-Obukhov Similarity Theory (MOST) utilized in Earth System Models. With slow trace gas sensors used in disjunct eddy-covariance (DEC) approaches and the more complex signal filtering associated with REA devices (here simulated using filtered signals from fast-response sensors), A22 outperforms REA and DEC in predicting the observed unfiltered (total) eddy-covariance (EC) fluxes. However, REA and DEC can still capture the observed filtered EC fluxes computed with the filtered scalar signal. This finding motivates the development of a correction, blending the REA and DEC methods, for the underestimated net averaged fluxes to incorporate the effect of sensor filtering. The only needed parameter for this correction is the mean velocity at the instrument height, a surrogate of the advective timescale.
\end{abstract}


\introduction  
The significance of surface-atmosphere exchanges of trace gases, volatile organic compounds (VOCs) and aerosol species to atmospheric composition and dynamics, and energy transport is not in dispute. Increasing concentrations of gases and particles due to natural and anthropogenic sources are modulating the Earth's climate and having deleterious consequences for human health and the environment \citep{qian2010investigation,kolb2010overview,voulgarakis2015interannual}. However, estimating these surface-atmosphere exchanges is particularly challenging in the stable atmospheric boundary layer (ABL) flows that are characterized by weak mixing and highly anisotropic turbulence \citep{stull1988introduction,mahrt1998nocturnal}. Stable ABLs occur at nighttime, in the downdraft region of deep mesoscale convective systems (that transport dry air from the mid troposphere to the surface where it is compressed to higher temperatures), and in polar regions; they persist as one of the least understood regimes in boundary layer meteorology owing to the inherently complex dynamics and the departure from continuous turbulence towards intermittency \citep{ansorge2014global,ansorge2016analyses,mahrt2020non,allouche2022detection}. On the sensing side, the so-called flux-gradient or flux-variance relations based on Monin-Obukhov Similarity Theory (MOST) are challenging to apply. This challenge is due to core assumptions that are tenuous to satisfy in practice for stable ABLs.  For example, a constant flux surface layer that requires stationarity, planar homogeneity, absence of subsidence, and a high Reynolds number state may not be well established for surface flux measurements under stable conditions. The challenges are exacerbated by surface heterogeneity, such as over surfaces with mixed water and sea ice in polar regions that can accelerate the exchanges of gases, aerosols, and energy between the ocean surface and the atmosphere \citep{sharma2012influence,fogarty2023atmospheric}, and semi-infinite heterogeneity patches e.g., land-sea interfaces \citep{allouche2023influence,allouche2023unsteady}. These observational challenges then propagate into theoretical and modeling considerations, prompting the need for improved estimates of scalar fluxes under stable conditions. 

To begin addressing these challenges and scientific gaps, turbulence and flux observations using the eddy covariance (EC) technique for fluxes of heat (an active scalar), momentum, and trace gases (representing passive scalars) are employed here.  These EC observations are then used to evaluate a series of models that can be employed to parameterize turbulent fluxes, either using slower and inexpensive sensors in field measurements through (i) the disjunct eddy-covariance (DEC) method and (ii) the relaxed eddy accumulation (REA) technique \citep{businger1990flux}, or (iii) using mean scalar concentrations that are available in coarse weather or climate models (mixing length-gradients models). In this study,
high-frequency measurements from two contrasting land-cover types are analyzed (i) over an ice sheet in Utqiagvik (Barrow), Alaska, and (ii) over a sparsely vegetated grassland downwind of heavy agriculture in Wendell, Idaho. Specifically, the current work seeks to answer the following research questions: (Q1) What flux/closure models can best reproduce the observed EC fluxes? Models that best describe the observed fluxes are then tested under scenarios that mimic coarse geophysical variables with mean fields measured using slow-response sensors because fast-response instruments remain largely unavailable for reactive chemical species (mainly those characterized by short atmospheric lifetimes). This motivates the second question: (Q2) Can the models correct for the "unresolved" turbulence scales inherently missed when data are collected using slow-response sensors? In this context, necessary modifications to the REA model are proposed to account for the filtering of the fast eddies associated with the REA device design. This then connects the analysis to the use of slow-response sensors in the DEC flux measurement approach. Since the scalar data are all available and collected using fast sensors, the slow response sensors used in the DEC method and the function of an REA device (the so-called dead-band) are both 'simulated' directly from time series of fast-response sensors as we will explain later.

\section{Theory}\label{Theory-models}

\subsection{Background and definitions}
Any instantaneous flow variable (e.g., $s$) is decomposed as $s=\overline{s}+s'$, where $\overline{s}$ is an “ensemble mean” quantity, and $s'$ is a turbulent quantity defined as a departure from $\overline{s}$. Operationally, primed variables are determined as excursions from the time-averaged state (hereafter indicated by the overbar).  The atmospheric stability is quantified using the dimensionless stability parameter $\zeta=z/L$, where $z$ is the wall-normal distance from the surface and $L$ is the Obukhov length \citep{obukhov1971turbulence}. Under stable conditions, which are the focus here, $\zeta>0$.

The strength of the variability in any flow variable is quantified by $\sigma_s=(\overline{s's'})^{1/2}$, the root-mean squared value of $s'$, while the covariance $\overline{w's'}$ is the average net vertical kinematic scalar flux with $w'$ being the vertical velocity fluctuation. From definitions, $\sigma_s$ is related to $\overline{w's'}$ using the correlation coefficient $R_{ws}$ defined as  
\begin{equation}\label{Model0}
R_{ws}=\frac{\overline{w's'}}{\sigma_w\sigma_s}.
\end{equation}

\subsection{VCC: Variable correlation coefficient flux model}
Using Eq. \ref{Model0}, a simplified flux model can be defined based on an empirical parametrization of $R_{ws}$ as a function of $\zeta$  
\begin{equation}\label{Model1}
\overline{w's'}=R_{ws}(\zeta)\sigma_w\sigma_s.
\end{equation}
The empirical relation ($R_{ws}(\zeta)$) here might still be non-generalizable as it may be site-specific and dependent on some other meteorological variables or surface conditions.

\subsection{ACC: Averaged correlation coefficient flux model}
Again using Eq. \ref{Model0}, one could also test another simplified model with an averaged correlation coefficient $\langle R_{ws} \rangle$, taken as the mean over all the available observational periods, yielding
\begin{equation}\label{Model2}
\overline{w's'}=\langle R_{ws} \rangle \sigma_w\sigma_s.
\end{equation}
In addition to assuming that the correlation coefficient is stability independent, the same potential drawbacks of the  variable correlation coefficient formulation also apply to this model, and the results could not be extrapolated to other sites where other factors may be present, such as heterogeneity, seasonality, and the influence of synoptic variability, to name a few.  

\subsection{REA: Relaxed eddy accumulation flux model}
\citet{businger1990flux} proposed the REA method to compute turbulent scalar fluxes. The REA method is ideally appropriate to use when fast-response sensors are available for $w'$ (typically from sonic anemometers) but only slow response measurements are available for the scalar concentration (slow trace gases sensors, or even trace gas samples that need to be collected and analyzed subsequently in a lab). Thus, the REA approach offers an enhanced representation of these scalar fluxes \citep{nie1995design}.

The basic idea here is inspired from the work of \cite{desjardins1977description}, who used conditional sampling techniques to collect scalar information (along with vertical wind speed) in two electronic counters, one for upflow and another for downflow. From linear correlation analysis, the regression slope of $w'/\sigma_w$ against $s'/\sigma_s$ may be estimated from the correlation coefficient following \citep{baker1992field,katul1994sensible,katul2018ejective}
\begin{equation}\label{Model2l}
R_{ws} = \frac{(\overline{s^+}-\overline{s^-})/\sigma_s}{(\overline{w^+}-\overline{w^-})/\sigma_w},
\end{equation}
where $\overline{s^+}$ is the conditional average of scalar $s$ instantaneously attributed to updraft events ($w'>0$), and $\overline{s^-}$ is the conditional average of scalar $s$ instantaneously attributed to downdraft events ($w'<0$), $\Delta s=\overline{s^+}-\overline{s^-}$ reflects this difference in collecting scalar information from the two samples, and likewise for the vertical velocity statistics. When this estimate for $R_{ws}$ is inserted into Eq. \ref{Model0}, the REA expression emerges as
\begin{equation}\label{Model2la}
 \overline{w's'} = \left[\frac{\sigma_w}{(\overline{w^+}-\overline{w^-})}\right] \sigma_w (\overline{s^+}-\overline{s^-}).
\end{equation}
For vertical velocity fluctuations that follow a Gaussian distribution, it can be shown that ${\sigma_w}/{(\overline{w^+}-\overline{w^-})}=\sqrt{2 \pi}/4$ \citep{katul2018ejective}, a constant whose numerical value is 0.63. 

Since the linear regression analysis to estimate $R_{ws}$ as featured in Eq. \ref{Model2l} is imperfect, an operational REA model can expressed as 
\begin{equation}\label{REA}
\overline{w's'}=\beta_s\sigma_w(\overline{s^+}-\overline{s^-}),
\end{equation}
where $\beta_s$ is now treated as an empirical coefficient that corrects for the above mentioned shortcomings.  Many studies investigated the choice of optimal $\beta_s$ over a wide range of stabilities, surfaces, and meteorological conditions \citep{businger1990flux,katul1996investigation,milne1999variation,zahn2016scalar,vogl2021choosing}. The choice of $\beta_s$ is still debatable, yet various studies reported a $\beta_s \approx 0.59$ \citep{bowling1998use,katul1996investigation}, which is not far from a Gaussian prediction derived from $w'$ statistics ($=0.63$). Hence, a $\beta_s=0.59$ is selected in the current study as a reference baseline in assessing the REA method.  

What is less debatable is the theoretical invariance of $\beta_s$ with stability changes: It was recently shown that the required independence of the REA formulation in the limit of free convection from the friction velocity ($u_*$) is not compatible with a stability dependent $\beta_s$ \citep{zahn2023relaxed}, and this stability invariance was in fact reported in many field observational studies. The arguments of \citet{zahn2023relaxed} for a stability-invariant $\beta_s$ can be deduced from the dimensionless form of the REA expression
\begin{equation}\label{REA_dim}
\frac{1}{\left[\beta_s\right]}=\frac{\sigma_w}{u_*}\frac{(\overline{s^+}-\overline{s^-})}{s_*},
\end{equation}
where $s_*=\overline{w's'}/{u_*}$. With $\overline{s^+}-\overline{s^-} \sim \sigma_s$, and noting that scalar flux-variance expressions of $\sigma_s/s_*$ exhibit opposite scaling exponents with $\zeta$ compared to $\sigma_w/u_*$ across all stability regimes, the dependence of $\beta_s$ on $\zeta$ is likely to be small as the two terms on the right hand side of Eq. \ref{REA_dim} cancel each others stability dependence.  In convective conditions, $\sigma_w/u_* \sim |\zeta|^{+1/3}$ whereas $\sigma_s/s_* \sim |\zeta|^{-1/3}$.  For near-neutral conditions, MOST predictions suggest $\sigma_w/u_* \sim |\zeta|^{0}$ and $\sigma_s/s_* \sim |\zeta|^{0}$ as well, making $\beta_s$ also independent of stability in that limit. 

Under stable conditions, similar plausibility arguments for a stability independent $\beta_s$ can be made based on the observations of \citet{Weaver1990} that $\sigma_s/s_* \sim |\zeta|^{0}$ also under very stable conditions. This result was explained by the author based on arguments first presented by \citet{Wyngaard1973} that under very stable conditions the active eddy size scales with $L$ rather than $z$, and thus turbulence statistics should become independent of $\zeta$. This would then also apply to $\sigma_w/u_*$, and by extension to $\beta$. However, as later shown in the present paper, (i) the practical application of REA using devices with finite mechanical response time to physically separate the accumulation of the trace gas in updrafts and downdrafts, (ii) the introduction of a `dead-band' at small $w'$  where the concentrations are not counted neither towards $\overline{s^+}$ nor towards $\overline{s^-}$, and (iii) the slow response of the scalar sensors may all induce an indirect stability dependence.

\subsection{DEC: Disjunct eddy covariance model}
The DEC technique is a close analogue of the classic eddy covariance technique, but here the scalar sensor has a slow physical response time. The sensor may still be sampled at a high rate, equal to that of $w^\prime$, to compute the DEC flux as $\overline{w^\prime s^\prime}$, but the user should be cognizant of the inherent filtering of the fluxes carried by eddies that the slow scalar sensor cannot resolve. In this paper, we simulate the filtered signal based on the actual high-frequency scalar concentration measurements; the details will be provided in Section \ref{filterdetails}.

\subsection{A22: Mixing length flux model}
Recently proposed models for momentum and heat fluxes based on mixing length analogies \citep{allouche2022detection} that outperformed MOST under stable periods marked with intermittent turbulence dynamics are also tested here. These models were initially formulated using an eddy diffusion representation of fluxes
\begin{equation}
\label{Eddy-Diffusivity}
\overline{w's'}=-K_s\frac{\partial\overline{s}}{\partial z}=-(\sigma_w L_{mix})\frac{\partial\overline{s}}{\partial z}.
\end{equation}
The eddy diffusivity ($K_{s}$) was then defined as the product of a characteristic velocity scale ($U_{char}$) and a mixing length scale ($L_{mix}$): $K_s=U_{char}L_{mix}$. Here, $L_{mix}$ will be defined differently for momentum ($K_u$) and heat ($K_{T_v}$, $T_{v}$ is the virtual temperature), but both use the standard deviation of the vertical velocity ($\sigma_w$) as the characteristic velocity scale (similar to REA), i.e., $K_s=\sigma_w L_{mix}$. 

For momentum, $L_{mix}=L_{u}$, and $L_{u}$ is defined as a harmonic average between two competing shear length scales ($L_{u1}$, a local turbulent shear scale, and $L_{u2}$, the classic bulk shear scale) as follows
\begin{subequations}\label{L-momentum}
\begin{align}
& L_{u1}=(1-\alpha_u)\sigma_w\left(\frac{\partial \overline{u}}{\partial z}\right)^{-1},\\
& L_{u2}=\alpha_u\overline{u}\left(\frac{\partial \overline{u}}{\partial z}\right)^{-1},\\
& L_{u}=\left( \frac{1}{L_{u1}} + \frac{1}{L_{u2}} \right)^{-1}.
\end{align}
\end{subequations}
In this model, $\alpha_u$ is an empirical constant; its value is determined as $\alpha_u=\alpha_{T_v}=0.35$ (same value found to be also adequate for the heat flux model described next). The mean wind speed at the measurement height is given by $\overline{u}$.

Similarly for heat, $L_{mix}=L_{T_v}$, and $L_{T_v}$ is defined as a harmonic average between two competing length scales.  The first is $L_{T_{v1}}$, the Ellison length scale \citep{ellison1957turbulent} and the second is $L_{T_{v2}}$, the buoyancy length scale \citep{stull1973inversion,zeman1977parameterization}. These scales are formulated as 
\begin{subequations}
\label{L-heat}
\begin{align}
& L_{T_{v1}}=\alpha_{T_v}\sigma_{T_v}\left(\frac{\partial \overline{T_v}}{\partial z}\right)^{-1}=\alpha_{T_v}\frac{\sqrt{2TPE}}{N_{BV}},\\
& L_{T_{v2}}=(1-\alpha_{T_v})\frac{\sigma_w}{N_{BV}},\\
& L_{T_v}=\left( \frac{1}{L_{T_{v1}}} + \frac{1}{L_{T_{v2}}} \right)^{-1}.
\end{align}
\end{subequations}
Here
\begin{equation}
N_{BV}=\sqrt{\left(\frac{g}{\overline {T_v}} \frac{\partial \overline{T_v}}{\partial z}\right)}
\label{eq:BV}
\end{equation}
is the Brunt-Väisälä frequency; $g$ is the gravitational acceleration; and $TPE$ is the turbulent potential energy, which is related to $N_{BV}$ as shown in $L_{T_{v1}}$ \citep{zilitinkevich2013hierarchy,katul2014two}. 


\subsection{MOST: Monin-Obukhov similarity theory flux model}
Based on dimensional analysis, \cite{MOST1954} formulated flux-gradient relations that are still used widely in weather prediction and climate models. MOST has inherent limitations as it applies to planar homogeneous conditions and stationary flows at very high Reynolds number in the absence of subsidence, and requires turbulent kinetic energy (TKE) production to be balanced by the TKE dissipation rate. Nevertheless, MOST still serves as reference for idealized conditions \citep{foken200650}. MOST fluxes could still capture the observed fluxes under weakly stable conditions where turbulence is continuously sustained and not intermittently suppressed.
MOST fluxes here are computed using the Businger–Dyer relations \citep{businger1971flux} as those relations remain pervasively in use today. Such relations are expressed by non-dimensional gradient (diabatic) functions, $\Psi_s(\zeta)$, relating the scalar concentration surface scale $s_*=\overline{w's'}/u_*$ to the gradient following
\begin{equation}
\label{MOST_Eq}
\Psi_s(\zeta)=\frac{\partial\overline{s}}{\partial z} \frac{{\kappa z}}{s_*}.
\end{equation}
where $\kappa$ is the von Kármán constant (= 0.4).
\section{Field Data and Methods}
In this study, data from two field experiments are analyzed. One data set is collected over the frozen tundra near Utqiaġvik, Alaska (U09) as part of the OASIS-2009 (Ocean-Atmosphere-Sea Ice-Snowpack) field campaign \citep{staebler2009role,perrie2012selected,bottenheim2013oasis}. The second data set is collected from November 2022 to January 2023 at a sparsely vegetated grassland in Wendell, Idaho (W22). At Utqiagvik, four sonic anemometers were mounted on a 10-m tall tower at 0.58, 1.8, 3.2, and 6.2 m above the snowpack, and the herein analyzed data correspond to $z_m$ = 1.8 m. Three-dimensional velocity ($u$,
$v$, $w$: longitudinal, lateral, and vertical components) and sonic virtual temperature ($T_s \approx T_v$, where $T_v$ is the true virtual temperature) measurements were recorded. 
At the Wendell site, data were acquired only at one height ($z_m$ = 2.4 m) above the ground surface. Chemical scalar concentration (carbon dioxide (CO$_{\text{2}}$), ammonia (NH$_{\text{3}}$), and water vapor (H$_{\text{2}}$O)), in addition to the three-dimensional velocity and temperature measurements, were recorded using a commercial open-path analyzer (CO$_{\text{2}}$/H$_{\text{2}}$O 7500A, LiCor Inc., Lincoln, NE), a custom made open path sensor with a quantum cascade laser (NH$_{\text{3}}$), and an R.M. Young 81000 sonic anemometer \citep{miller2014open,sun2015open,pan2021ammonia}. Both gas sensors were subject to the necessary spectroscopic corrections arising from temperature and density fluctuations. Precision-time-protocol (PTP) was used to ensure all the gas, environmental, and meteorological sensors were synchronized by GPS (Global Positioning System). For both sites, instantaneous molar density measurements of the chemical species were converted to mass concentrations using the pressure sensor on the LiCor 7500A, which has an accuracy of ±0.4 kPa from 50 to 110 kPa and a resolution of 0.006 kPa \citep{edson2011direct}. The gas fluxes were then calculated based on their mass concentrations, in lieu of applying the so-called WPL density corrections to the fluxes after processing \citep{webb1980correction,detto2007simplified}.

The sampling frequency at both sites was set at $f_s=10$ Hz, and the post-processing involved (i) de-spiking, (ii) linear detrending, and (iii) double rotation of wind components \citep{wilczak2001sonic}. Fluxes and other required statistics were then computed for various scalar quantities (i.e.,  $T_v$, CO$_{\text{2}}$, NH$_{\text{3}}$, and H$_{\text{2}}$O; $u$ and its associated momentum flux were also tested here for comparison). Analysis periods were set to 15-min in U09 (the 15-min Reynolds average choice here is selected because U09 periods reveal strong intermittent behaviour) and 30-min W22; these were then the periods used for double rotation and time-averaging throughout. Details of the data quality control for these data sets can be found elsewhere \citep{allouche2022detection}.

\section{Results and discussion}
An earlier study \citep{zahn2023relaxed} investigated REA under non-ideal unstable conditions and concluded that the REA method outperforms MOST flux models. One of the main aims of the present study is to examine whether REA outperforms MOST under stably stratified conditions. MOST is used as a reference for comparison as it reflects the 'state-of-the science' in climate models. Since A22 established the limitations of MOST under stable conditions for the Utqiagvik data set and further proposed the closure models detailed previously that also outperform MOST, REA will then be compared to A22. Other model details in Section \ref{Theory-models} will serve as additional benchmarks to understand model performance, but the analyses focus on the REA and A22, and an REA-DEC hybrid approach. 

\subsection{Model inter-comparison using high frequency measurements}
\begin{figure*}[h!]
\includegraphics[width=18cm]{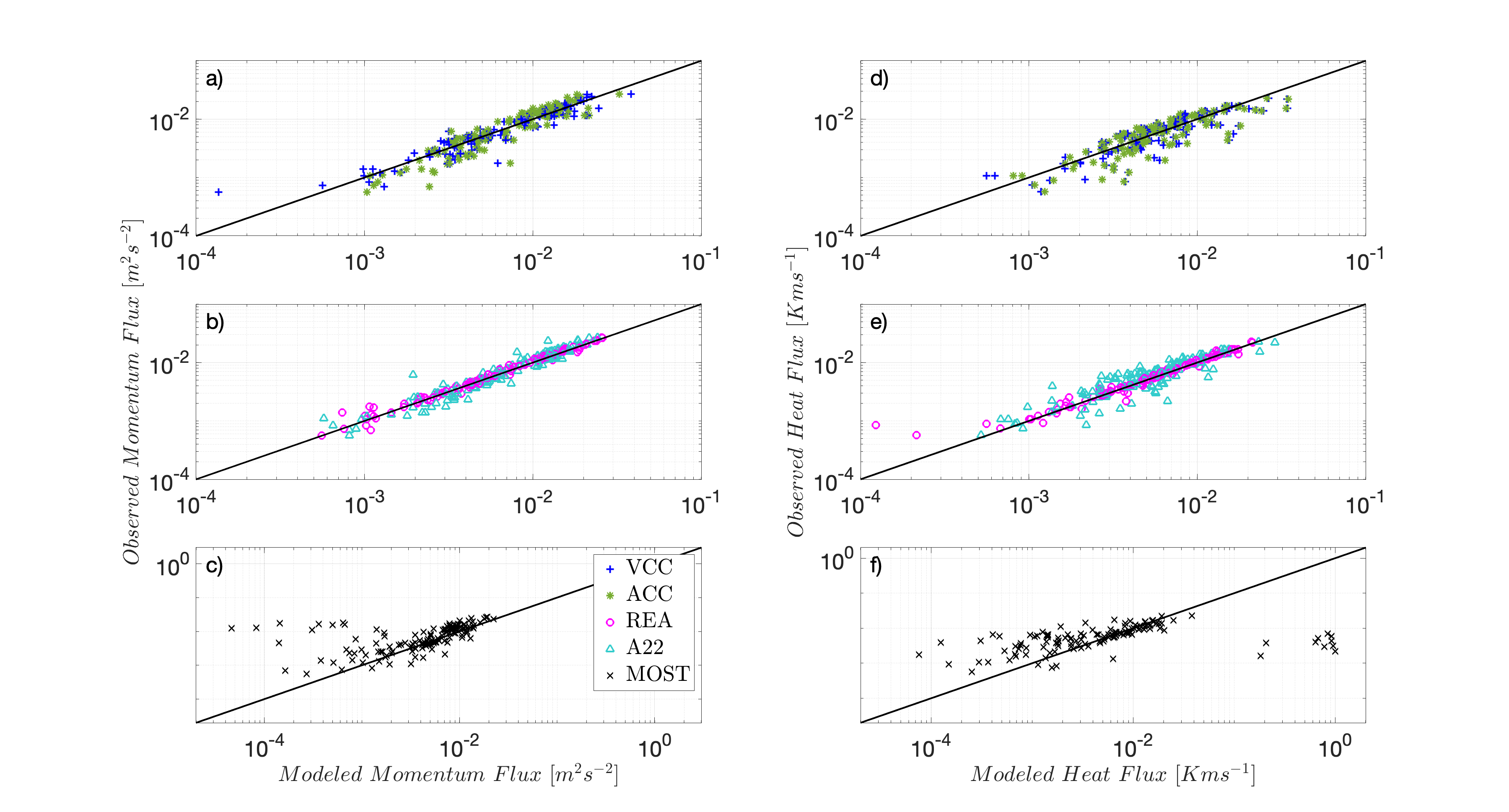}
\caption{(a,b,c) Inter-comparison of kinematic momentum fluxes and (d,e,f) kinematic heat fluxes derived from the various models  for the Utqiagvik site U09. One-to-one line is shown as a reference (solid black line). Since both fluxes are negative, they are multiplied by $-1$ to plot on log-log scale.}
\label{fig:1}
\end{figure*}

All introduced models are now tested at the Utqiagvik site because it has the multiple levels that are required for testing the A22 and MOST models. The middle panel subplots of Fig. \ref{fig:1}, corresponding to the REA and A22 models, depict the strongest correlation between modeled and observed EC fluxes for both momentum and heat. In addition, the ACC model, which incorporates stability information (Eq. \ref{Model1}), performs slightly better than the constant ACC model (Eq. \ref{Model2}). Although these models at first may appear to be better approximations of the eddy-covariance fluxes than REA, the REA model performance is in practice superior, benefiting from the cancellation of the effect of stability in the model coefficient $\beta_s$ as detailed previously. This agrees with prior findings \citep{zahn2023relaxed} for unstable conditions. All proposed models outperform MOST (Fig. \ref{fig:1}c-\ref{fig:1}f). With their superior performance established, the REA and A22 models will be the only ones retained in the subsequent analyses.

\subsection{Simulating a slow scalar sensor for model testing}\label{filterdetails}
To address the limited bandwidth of many trace gases sensors, we simulate the output of a real slow sensor $\widetilde{s}$ measuring a variable $s$ as the numerical solution to the first-order ODE in Eq. \ref{ODE}. Here, $s$ would be the "fast" turbulent sensor, and $\overline{\Delta}$ is the time scale of the filter-width (the response time scale of the slow sensor), which we selected to vary in the range [0-5 s] with increments of 1 s. Thus, solving the following equation numerically (using explicit forward Euler time advancement scheme) converts the $s$ time series from 10-Hz to a lower frequency, down to 0.2-Hz when $\overline{\Delta}$ = 5 s:
\begin{equation}
\label{ODE}
\frac{d\widetilde{s}}{dt}+\frac{1}{\overline{\Delta}}\widetilde{s}= \frac{1}{\overline{\Delta}}s.
\end{equation}
The observed filtered EC fluxes computed with the filtered scalar signal $\widetilde{s}$ mimic the fluxes estimates of the disjunct eddy-covariance method, (DEC) fluxes \citep{ruppert20022}. We also tested another filter-type, a low-pass Gaussian filter, and results were not sensitive to the filter type, but the ODE solution is a more accurate model for a first-order slow sensing system. The filtering is not applied to the vertical velocity ($w$) as high-frequency anemometers are readily available. Hence, a tilde denotes the filtered virtual temperature ($\widetilde{T_v}$), scalar concentration ($\tilde s$), and streamwise velocity ($\widetilde{u}$). All three-dimensional velocity components ($u$, $v$, and $w$) are available in high-frequency output of the sonic anemometer, but only $u$ is filtered ($\widetilde{u}$) to compare its kinematic momentum fluxes to those of scalars. 

In theory, an REA system should not suffer from the slow response of the trace gas sensor since it only requires the mean measurements of $\overline{s^+}$ and $\overline{s^-}$. However, this would require a mechanism to separate the gas streams that has a 10-Hz response time as well. Given the limited responsive feedback time of the opening and closing mechanisms of the valves in real REA systems and to avoid an excessive number of movements, a certain threshold value of a dead-band velocity $w_0$ is selected to guarantee larger individual air samples. A dynamic dead-band that is linked to the turbulence conditions in each period is adopted such that similar amounts of air are sampled in updrafts and downdrafts. We used the empirical finding of \cite{grelle2021affordable}, as depicted in Eq. \ref{w_0}, which yielded roughly similar amounts of air in the updraft and downdraft reservoirs. Sampling is activated only if the vertical wind speed exceeds this threshold value in $w_0$ i.e., $\lvert w \rvert >w_0$.
\begin{equation}
\label{w_0}
w_0=\frac{\sigma_w}{3.5}.
\end{equation}
We should underline that the computations of $\overline{s^+}$ and $\overline{s^-}$ for the REA are first done with the unfiltered $s$ signal, and the dead-band of vertical velocity defined above is then the only effective filter that applies to the REA computations. However, to examine possible mechanical filtering and explore an REA-DEC hybrid approach that we will later detail, we also compute the REA fluxes with $\overline{s^+}$ and $\overline{s^-}$ from the filtered signal produced by Eq. \ref{ODE}.

\subsection{DEC, REA and A22 model evaluation using simulated slow sensor data}
\begin{figure*}[t]
\includegraphics[width=18cm]{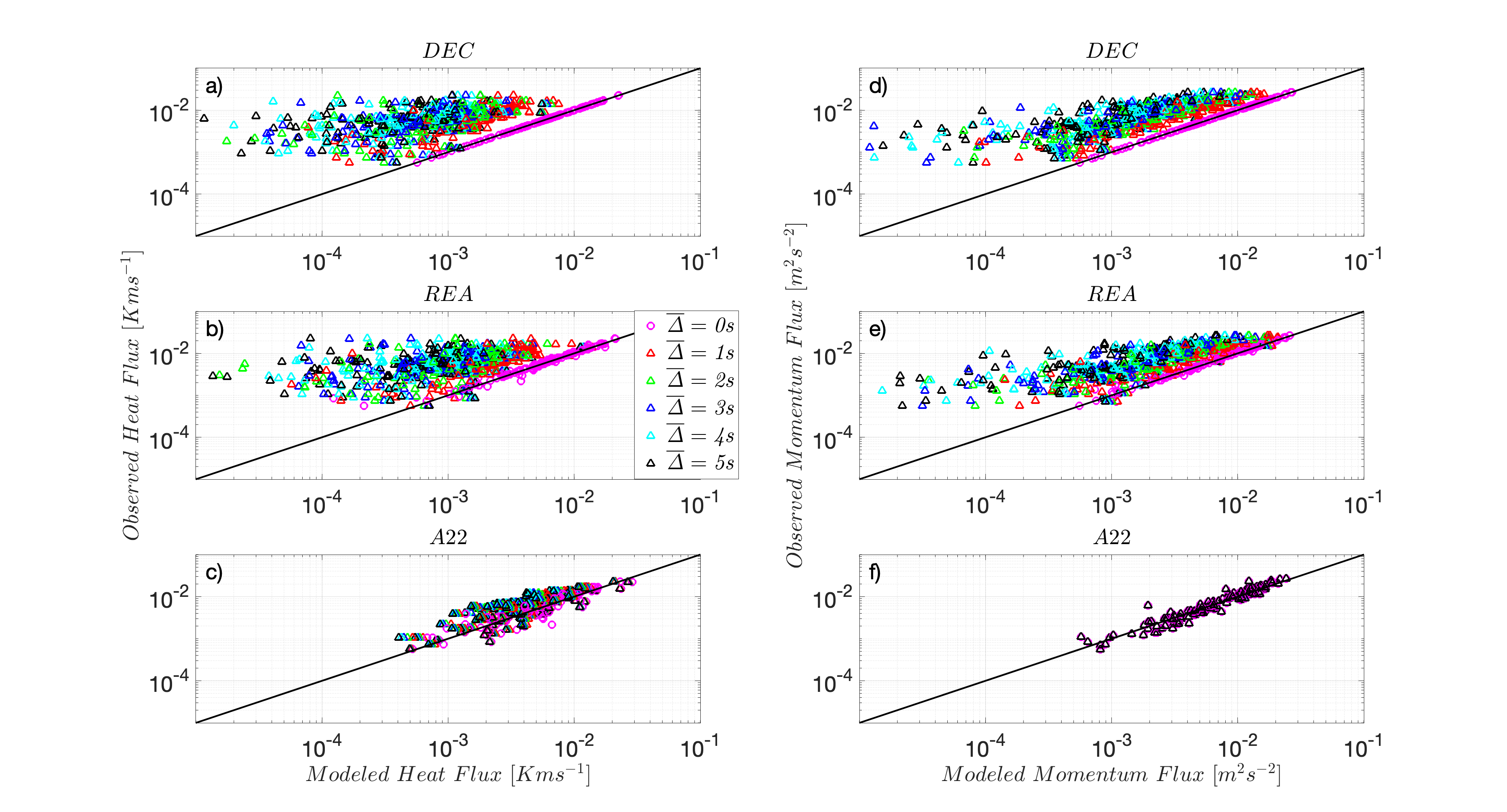}
\caption{Heat flux using $\widetilde{T_v}$ predicted by DEC (a), REA (b) and A22 (c) models, and momentum flux using $\widetilde{u}$ predicted by DEC (d), REA (e) and A22 (f) models at the Utqiagvik site U09, compared to real observed EC fluxes. One-to-one line is shown as a reference (solid black line). Heat fluxes are multiplied by $-1$ to plot on log-log scale. $\overline{\Delta}$ (s) is the filter time scale.}
\label{fig:2}
\end{figure*}
Again focusing on Utqiagvik data set with its multilevel measurements, the models are now tested using inputs from a slow sensor. The top and middle panel subplots of Fig. \ref{fig:2} show that DEC and the REA methods significantly, and in similar fashion, underestimate the observed (unfiltered) heat and momentum EC fluxes based on filtered quantities $\widetilde{T_v}$ and $\widetilde{u}$, respectively, as $\overline{\Delta}$ increases. This underestimation, however, is not surprising because under stable conditions small eddies carry a significant proportion of the fluxes, especially when the background flow is laminarizing and intermittent as shown elsewhere \citep{ansorge2014global,ansorge2016analyses,allouche2022detection,issaev2023intermittency}. These small eddies are filtered appreciably by the sensor's slow response. We should underline here that an REA system with fast-response mechanical valves will give results equivalent to the unfiltered REA ($\overline{\Delta} = 0$), and will be in good agreement with the actual EC fluxes. However, it is more likely that various REA devices will introduce some type of filtering to the signal that depends on the device design. Interestingly, the bottom panel subplots of Fig. \ref{fig:2}c and f show that the A22 model's performance is not sensitive to the signal filtering and provides good estimates of the observed (unfiltered) heat and momentum EC fluxes, even when filtered quantities $\widetilde{T_v}$ and $\widetilde{u}$ are used, and up to the highest filter-width $\overline{\Delta}$ = 5 s. The A22 model relies on multilevel means and variances in computing the fluxes, which tend to be carried by larger scale than the actual fluxes; this may explain the independence of the model performance from sensor response.

\begin{figure*}[t]
\includegraphics[width=18cm]{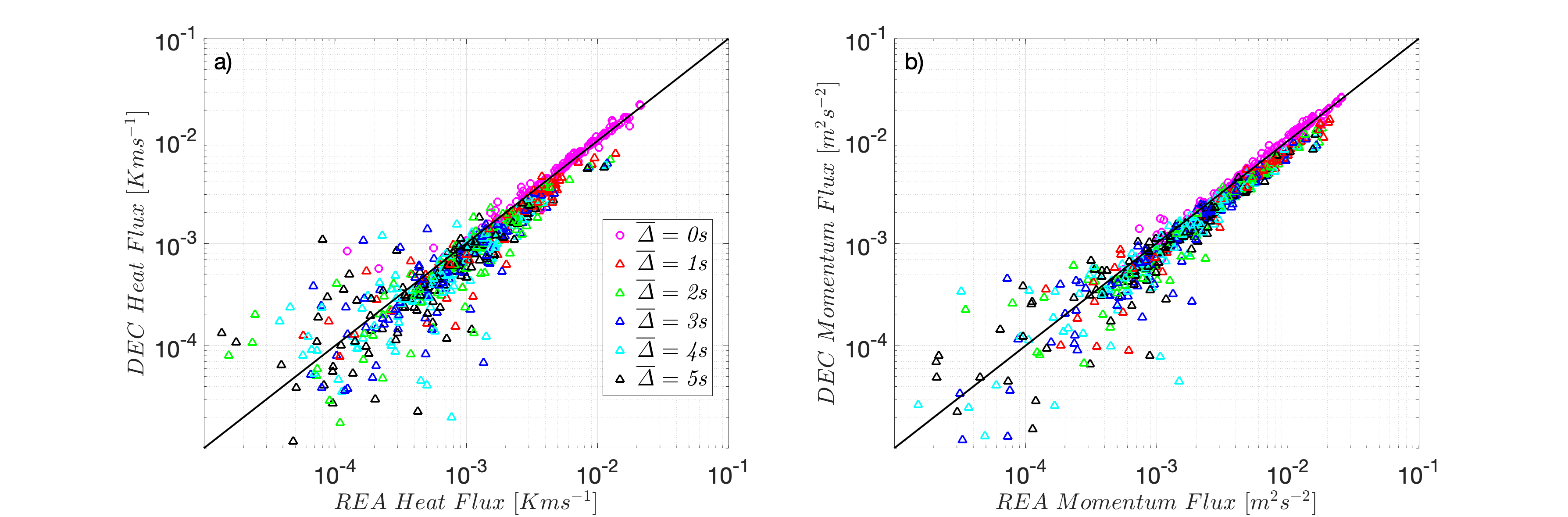}
\caption{(a) Heat flux using $\widetilde{T_v}$ predicted by REA model versus the filtered observed heat DEC flux and (b) momentum flux using $\widetilde{u}$ predicted by REA model versus the filtered observed momentum DEC flux, at the Utqiagvik site U09. One-to-one line is shown as a reference (solid black line). Heat fluxes are multiplied by $-1$ to plot on log-log scale. $\overline{\Delta}$ is the time scale filter-width (s).}
\label{fig:3}
\end{figure*}

Given the results above, a followup question is whether the REA model estimates with a filtered signal actually correspond to the fluxes that would be computed using eddy covariances of the filtered scalar signal, i.e., the DEC fluxes. Fig. \ref{fig:3}a and b indeed show that the REA model is still a reliable method to capture these filtered observed heat and momentum DEC fluxes at Utqiagvik as $\overline{\Delta}$ increases. All quantities here, including the fluxes, are computed based on filtered series $\widetilde{T_v}$ and $\widetilde{u}$. This implies that a filtering of the REA signal by slow mechanical devices is broadly comparable to the filtering of the DEC fluxes by slow-response scalar sensors.

\begin{figure*}[t]
\includegraphics[width=18cm]{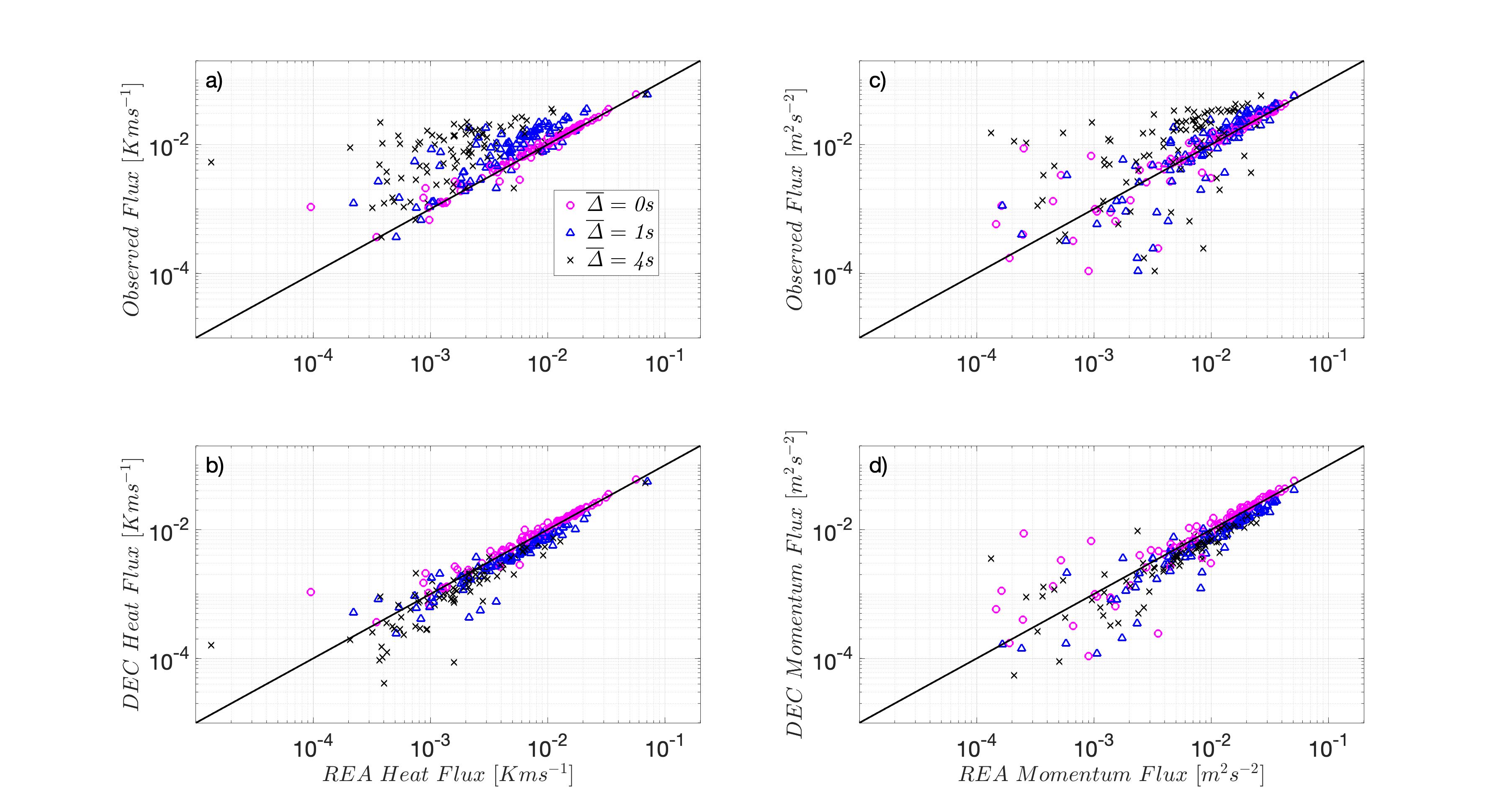}
\caption{Heat flux using $\widetilde{T_v}$ predicted by REA versus: the observed unfiltered EC flux (a) and the filtered DEC flux (b), and momentum flux using $\widetilde{u}$ predicted by REA versus: the observed unfiltered EC flux (c) and filtered DEC flux (d), at the Wendell site W22. One-to-one line is shown as a reference (solid black line). Heat fluxes are multiplied by $-1$ to plot on log-log scale. $\overline{\Delta}$ is the time scale filter-width (s).}
\label{fig:4}
\end{figure*}

\begin{figure*}[t]
\includegraphics[width=18cm]{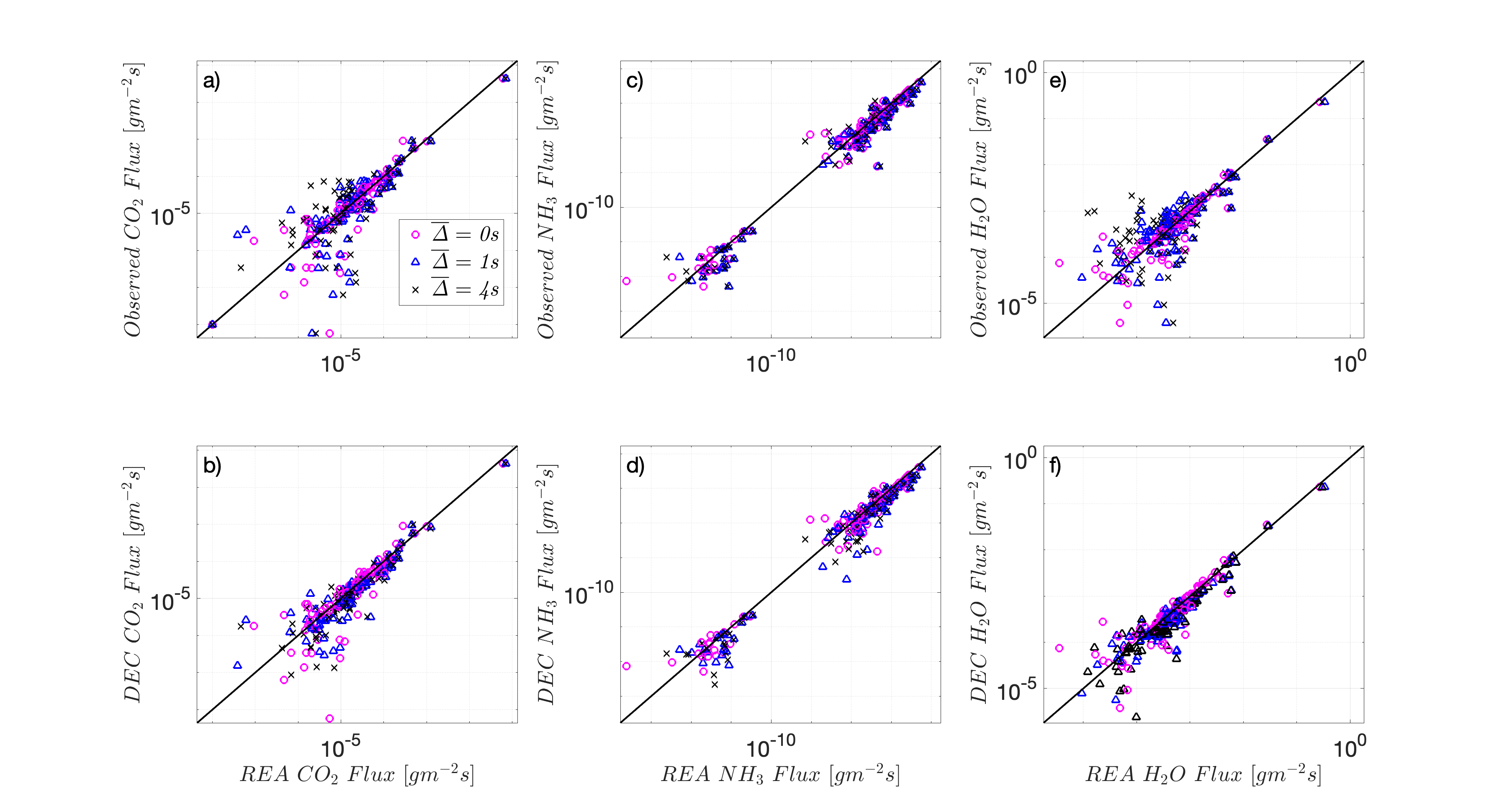}
\caption{ Similar to Fig. \ref{fig:4}, but here for scalars, (a,b): CO$_{\text{2}}$, (c,d): NH$_{\text{3}}$, and (e,f): H$_{\text{2}}$O.}
\label{fig:5}
\end{figure*}

This match between REA and filtered DEC fluxes is also observed at the the Wendell site for heat and momentum, as well as for the other scalars available at that site (CO$_{\text{2}}$, NH$_{\text{3}}$, and H$_{\text{2}}$O). As in Utqiagvik's site, the bottom panels of Figs. \ref{fig:4} and \ref{fig:5} show that REA performs better when evaluated against the filtered DEC fluxes, compared to the respective top panels of Figs. \ref{fig:4} and \ref{fig:5} that compare it to total EC fluxes. Note that since one level of measurements at Wendell is available, gradients of first-order moments could not be computed, which precludes testing of MOST or A22 models.

These common REA findings among the two different sites indicate that, under stable conditions, the REA captures the fluxes of the `resolved' eddies. Any mechanical response filtering will cause the method to significantly underestimate the needed high-frequency, full observed fluxes. An important question (addressed in the next subsection) that follows is whether the REA fluxes can be `corrected' under stable conditions to recover the missed fluxes. The proposed modifications, outlined below for the REA methodological framework, will also offer a correction for DEC fluxes to recover their corresponding total EC fluxes. It is to be noted that under unstable conditions at the Wendell site, where flux carrying eddies are of much larger scales than under stable conditions, the REA method is found to be almost insensitive to the considered filter-widths ($\overline{\Delta}$'s) (refer to Appendix \ref{Appendix:A}).  Therefore, REA performs well and captures the observed (high and low frequency) fluxes under unstable convective regimes, and hence biases in predicting scalar fluxes are expected to be minimal (Figs. \ref{fig:7} and \ref{fig:8}). An effective correction for the under-resolved fluxes under stable conditions will thus provide a method to obtain continuous accurate fluxes using REA over the whole diurnal cycle.

\subsection{A sensor-response correction for the optimal REA coefficient $\beta_s$}
Fig. \ref{fig:3} reveals that the scatter between REA and EC or DEC results is larger when the fluxes are small, which would translate into larger errors and scatter in the values of $\beta_s$. Analyses not shown here also reveal that $\beta_s$ values calculated for each period converge well towards the 0.59 value when the correlation $R_{ws}$ increases i.e., $R_{ws}$ > 0.2, indicating larger fluxes, with more scatter for lower correlation values. However this scatter is randomly distributed around the 1:1 line, indicating some error cancellation when the fluxes are integrated over longer periods of time. Discussions on these random variations in $\beta_s$ have linked them to the effect of height above canopies \citep{gao1995vertical}, and to the energy content influence of the associated eddy motions \citep{katul1996investigation}, among others, and are not a focus of the present paper.

Figs. \ref{fig:2} and \ref{fig:4} top panel, on the other hand, reveal that the chosen value of $\beta_s$ = 0.59 in the modeled fluxes, is a good estimate in recovering the observed EC fluxes when the signal is not filtered ($\it{\overline{\Delta} = \rm 0s}$). Missing smaller eddies (when the signal is filtered) that contribute significantly to fluxes under stable, but not unstable, conditions thus requires a larger $\beta_s$ to predict the correct fluxes using REA. Such underestimation was attributed to the filtering operation, dictated here by the choice of $\overline{\Delta}$, although we must underline that in actual REA the filtering is dictated by the mechanical system and its response time. This motivates a model development for $\beta_s$ that incorporates the effect of filtering. 

However, the agreement between the DEC and REA predictions when the scalar signal is filtered also opens the possibility to apply the REA method without a device that separates air streams from downdrafts and updrafts. If a slow response sensor (open or closed path) akin to those used in DEC is available, the REA equation can be applied with $\overline{s^+}$ and $\overline{s^-}$ computed using conditional averaging of the scalar time series based on the sign of $w$, like we are doing in this study with the simulated REA measurements, but without the dead-band given in Eq. \ref{w_0}. This approach would also then require a correction to $\beta_s$; we will hereafter refer to this method as the REA-DEC hybrid.

\begin{figure*}[t]
\includegraphics[width=18cm]{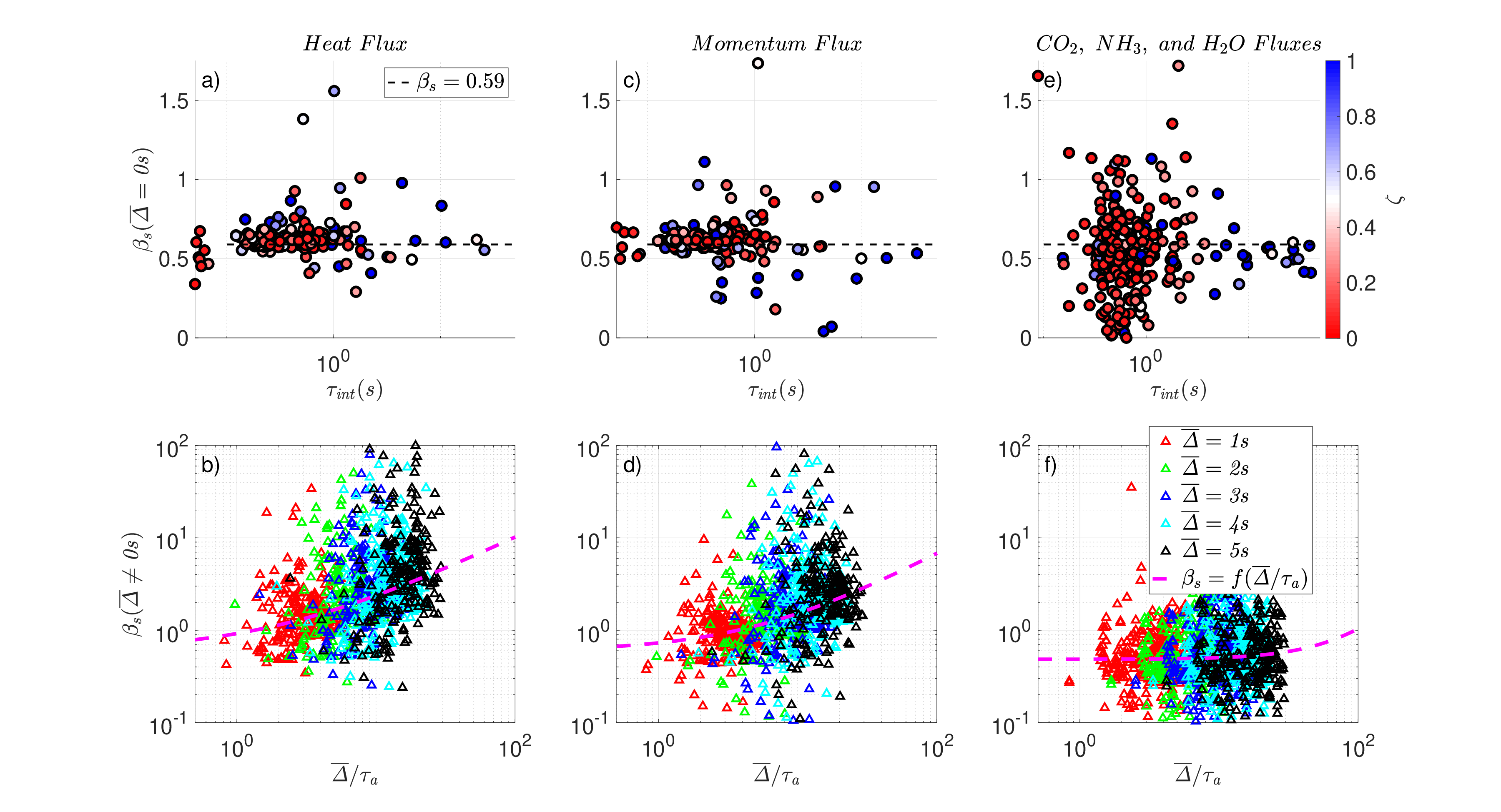}
\caption{ (a,c,e): Scatter of the computed $\beta_s$ for the observed (unfiltered) heat, momentum and passive scalar (CO$_{\text{2}}$, NH$_{\text{3}}$, and H$_{\text{2}}$O) fluxes at both sites relative to the integral time scale $\tau_{int}$(s), respectively. (b,d,f): similarly the computed $\beta_s$ for the respectively observed (filtered) fluxes relative to $\overline{\Delta}/\tau_{a}$. $\tau_{a}$ is an advective time scale, and the magenta solid lines refer to the empirical fit models $\beta_s=a\left(\overline{\Delta}/\tau_{a} \right)^{b}+c$, refer to Table \ref{table:1}.}
\label{fig:6}
\end{figure*}

For this purpose, we first compute each period's optimal $\beta_s$ (i.e., by inverting expression \eqref{REA}) that causes the REA predicted fluxes to match the exact observed fluxes when $\it{\overline{\Delta} = \rm 0s}$ (no filtering) across the two contrasting sites. The top panel subplots, Fig. \ref{fig:6}a  for heat, Fig. \ref{fig:6}c for momentum, and Fig. \ref{fig:6}e for all three passive trace gases (CO$_{\text{2}}$, NH$_{\text{3}}$, and H$_{\text{2}}$O), show a scatter plot of these exact $\beta_s$'s relative the integral time scale $\tau_{int}$(s) for each period's co-spectrum, where the data points are colored with the MOST stability parameter. The integral time scale ($\tau_{int}$) is determined by integrating the 
autocorrelation function ($\rho_{{w's'}}(\tau)$) up to the first zero-crossing from the measured ${w's'}$ instantaneous time series. We observe  $\beta_s$ values that depart from $\beta_s=0.59$ under all stabilities, but in general there is no clear $\beta_s$ dependence on $\tau_{int}$ as depicted here. 

For $\it{\overline{\Delta} \neq 0s}$, the corresponding bottom panel subplots, Fig. \ref{fig:6} (b,d,f), show these same $\beta_s$'s relative ${\overline{\Delta}}/\tau_{a}$ for different $\overline{\Delta}$'s, where $\tau_{a}$ is an advective time scale. After experimenting with various choices of time scales for normalizing the filter scale $\overline{\Delta}$, an advective time scale formed by $z$ and $\overline{u}$, hereafter labeled as $\tau_{a}=(\kappa z/\overline{u})$, appears to provide the best scaling for the variations of $\beta_s$ with the filter size ($\kappa$ is the von Kármán constant)
. This converges with the work of \citet{horst1997simple} who formulated corrections to estimate the attenuation of scalar flux measurements by slow response of sensors (akin to DEC). They used the sensor response frequency and a normalized frequency formulated based on $\tau_a$, which is the frequency of the peak of the logarithmic cospectrum $f_m=n_m \kappa / \tau_a$ to correct for the missed fluxes. In that work, $n_m$ is the dimensionless frequency at the cospectral maximum where it is estimated from observations of its behavior as a function of atmospheric stability $\zeta$. The advective time scale was similarly found to be a plausible choice in describing the drift and non-linear diffusion terms of a proposed non-linear Langevin equation to model the turbulent kinetic energy in stably stratified ABL \citep{allouche2021probability}. This characteristic time scale, $\tau_{a}$, measures the advection time of the attached eddies, of size $\kappa z$, past a fixed sensor.  

As depicted in Fig. \ref{fig:6}, an empirical fit that relates $\beta_s$ to $\overline{\Delta}/\tau_{a}$ i.e., $\beta_s=f\left(\overline{\Delta}/\tau_{a} \right)$ is proposed here to recover the real observed (unfiltered) EC fluxes using the REA method with ($\widetilde{s}$) measurements, either due to physical filtering by the device or due to the use of the REA-DEC hybrid approach proposed above. This relation is best described using a power-law model, $\beta_s=a\left(\overline{\Delta}/\tau_{a} \right)^{b}+c$, and Table \ref{table:1} summarizes these empirical coefficients $(a,b)$ for momentum, heat (active scalar), and passive scalar (CO$_{\text{2}}$, NH$_{\text{3}}$, and H$_{\text{2}}$O) fluxes at both sites where $c=\beta_s(\overline{\Delta}=0)$. The reported $b$ exponent for all fluxes, which describes how $\beta_s$ scales with $\overline{\Delta}/\tau_{a}$, varies between active and passive scalars. Note that if the dead-band criterion is removed, $b \approx 0.7$ and does not vary much among all scalars as shown in Table \ref{table:2}, hinting at the possible universality of such dependence for an REA-DEC hybrid model that removes the need for a complex mechanical REA system. Nevertheless, further exploration at disparate sites, and analyses of observational data for different scalars across wider stability ranges, are needed to have increased statistical confidence in the reported values of $(a,b)$ and their generalizability, especially for the Utqiagvik site that does not have measurements of the passive scalars. 

A common feature for all fluxes, as depicted in the bottom panel of Fig. \ref{fig:6}, is that the proposed model becomes less certain as $\overline{\Delta}$ increases (as expected). Therefore, such a model becomes less reliable with slow sensors that cannot resolve the integral statistics of turbulent flow variables. Otherwise, to compute $\beta_s$, the only needed inputs are (i) $\overline{\Delta}$ (provided by the slow sensor manufacturer, or it must be computed for a given REA device) and (ii) $\tau_{a}$ (computed from mean wind measurements). The reported fitting parameters in Table \ref{table:1} are obtained using the least absolute residuals (LAR) method, so that extreme values, which occur less frequently and may be related to unusual conditions or measurement errors, have a lesser influence on the fit. Given the variability in the optimal values of $\beta_s$, the proposed models would be mostly suited for quantification of long-term aggregates (net averaged fluxes) of the scalars cycle.

\begin{table*}[t]
\caption{Proposed models incorporating both sites Utqiagvik (U09) and Wendell (W22) with accounting for the dead-band criterion, $\beta_s=a\left(\frac{\overline{\Delta}}{\tau_{a}} \right)^{b}+c, \ c=\beta_s(\overline{\Delta}=0)$}
\begin{tabularx}{0.85\textwidth} { 
  | >{\centering\arraybackslash}X 
  | >{\centering\arraybackslash}X 
  | >{\centering\arraybackslash}X 
  | >{\centering\arraybackslash}X 
  | >{\centering\arraybackslash}X 
  | >{\centering\arraybackslash}X | }
 \hline
   Fitting Parameters $\beta_s=a\left(\frac{\overline{\Delta}}{\tau_{a}} \right)^{b}+c$ & Heat \  (active scalar) & Momentum & CO$_{\text{2}}$, NH$_{\text{3}}$, and H$_{\text{2}}$O \ (passive scalars)\\
   \hline
   $a$ & $0.3309$ & $0.1391$ & $0.0007$ \\
  \hline
   $b$ & $0.7316$ & $0.825$ & $1.448$\\
   \hline
   $c$ & $0.59$ & $0.59$ & $0.48$\\
\hline
\end{tabularx}
\belowtable{} 
\label{table:1}
\end{table*}
\begin{table*}[t]
\caption{Proposed models incorporating both sites Utqiagvik (U09) and Wendell (W22) without accounting for the dead-band criterion, $\beta_s=a\left(\frac{\overline{\Delta}}{\tau_{a}} \right)^{b}+c, \ c=\beta_s(\overline{\Delta}=0)$}
\begin{tabularx}{0.85\textwidth} { 
  | >{\centering\arraybackslash}X 
  | >{\centering\arraybackslash}X 
  | >{\centering\arraybackslash}X 
  | >{\centering\arraybackslash}X 
  | >{\centering\arraybackslash}X 
  | >{\centering\arraybackslash}X | }
 \hline
   Fitting Parameters $\beta_s=a\left(\frac{\overline{\Delta}}{\tau_{a}} \right)^{b}+c$ & Heat \  (active scalar) & Momentum & CO$_{\text{2}}$, NH$_{\text{3}}$, and H$_{\text{2}}$O \ (passive scalars)\\
   \hline
   $a$ & $0.1456$ & $0.0835$ & $0.0096$ \\
  \hline
   $b$ & $0.7309$ & $0.7169$ & $0.7098$\\
   \hline
   $c$ & $0.59$ & $0.59$ & $0.59$\\
\hline
\end{tabularx}
\belowtable{} 
\label{table:2}
\end{table*}

\conclusions  
Conventional and novel closure models are assessed in this study, with an emphasis on the REA method, to predict scalar fluxes under stable conditions. The models were tested using measurements collected at two different sites (Utqiagvik and Wendell). It was found that the REA and A22 models outperform the conventional models (e.g., MOST), and are thus less sensitive to departures from ideal flow conditions of homogeneity, steadiness, negligible vertical transport, and TKE production-dissipation balance. The A22 model was found to be insensitive to a filtering of the turbulent scales because it only requires the means at different heights. The REA model, on the other hand, is sensitive to any filtering that would be induced by a slow response of its mechanical components needed to separate the air streams from the updrafts and downdrafts. It is much less affected by the $w_0$ dead-band imposed to avoid too many operations of the air sampling valves if the latter are rapid.

In numerically simulating slow sensors, it was noted that the A22 model outperforms REA in predicting the observed (unfiltered) EC fluxes; however, REA can still capture the filtered observed DEC fluxes. This suggests that an REA approach can be implemented without a physical device to separate the updraft and downdraft air streams. Using a DEC-like slow sensor, one can conditionally average the concentrations using the sign of $w'$. To correct the underestimated REA fluxes in such an REA-DEC hybrid approach or due to physical device filtering, relative to the observed (unfiltered) EC fluxes, a model for the $\beta_s$ factors in the conditional sampling that incorporates the effect of filtering is proposed. The relation found to best describe this effect is a power-law model given by $\beta_s=a\left(\overline{\Delta}/\tau_{a} \right)^{b}+c$. The reported empirical coefficients $(a,b,c)$ for heat, momentum, and the passive scalars (CO$_{\text{2}}$, NH$_{\text{3}}$, and H$_{\text{2}}$O) fluxes in Table \ref{table:1} still need to be tested over a wider range of non-ideal surfaces and atmospheric conditions. The corrected REA model is less certain with extremely slow mechanical systems or sensors (i.e., as $\overline{\Delta}$ increases), yet it remains robust in terms of reproducing long-term averages of the scalar fluxes across their ecosystem lifetime cycle. Use of this model, along with the observations reported in Appendix \ref{Appendix:A} that reveal an insensitivity of $\beta_s$ to sensor response under unstable regimes, suggest that REA can be a robust framework for estimating turbulent fluxes when only single level measurements are available, but with sensors that can still resolve the integral scales of turbulence. The A22 can be alternatively used, under stable conditions, when multilevel measurements are available to compute needed mean gradients, even with slow-response sensors.



\dataavailability{The dataset of all the observational data for the two field experiments (Barrow and Wendell) are publicly available at (https://doi.org/10.5281/zenodo.10073726).} 




\appendix
\section{REA performance under unstable conditions at Wendell}\label{Appendix:A}    
\begin{figure*}[t]
\includegraphics[width=18cm]{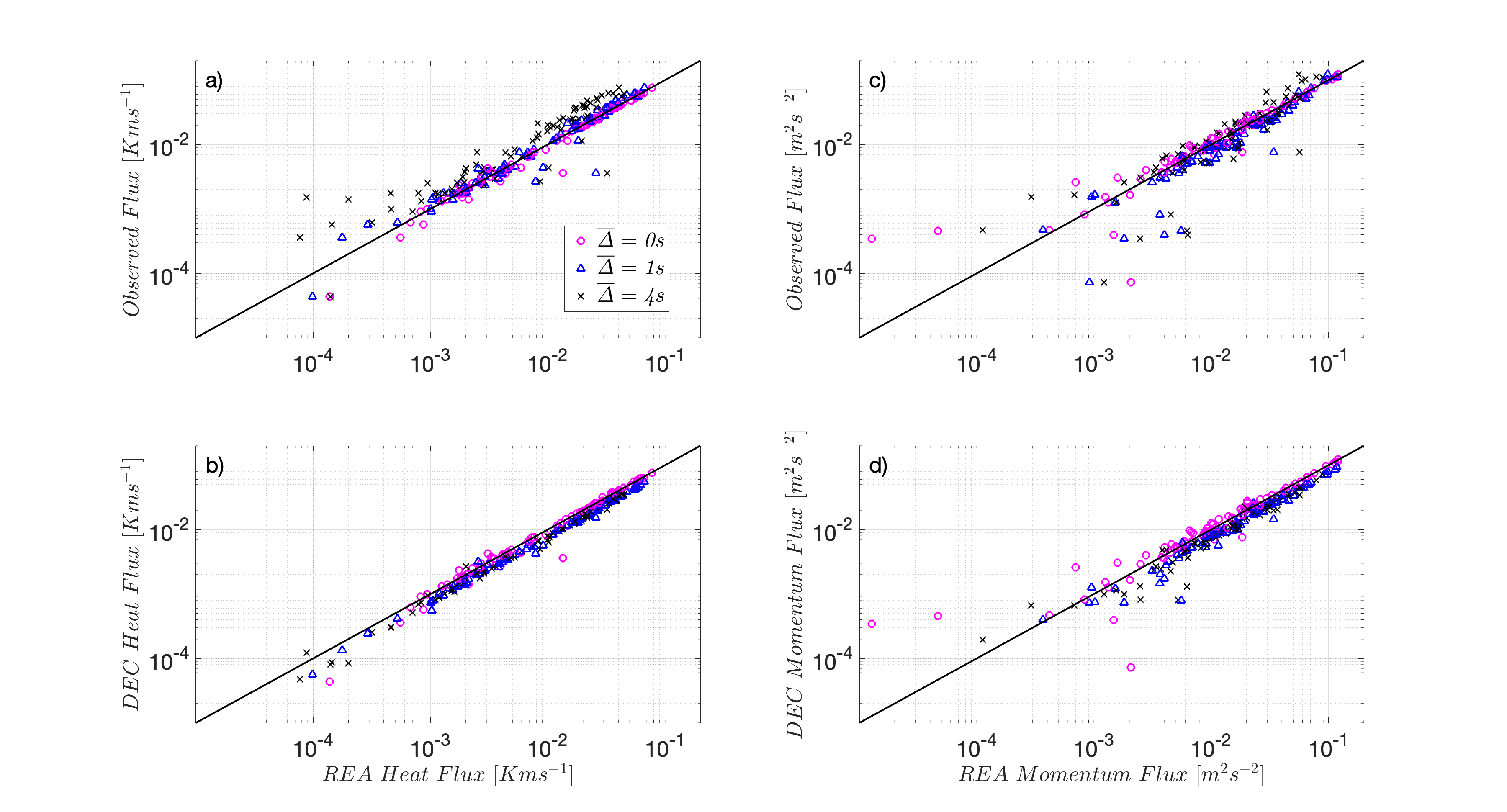}
\caption{ Similar to Fig. \ref{fig:4} but under unstable conditions.}
\label{fig:7}
\end{figure*}

\begin{figure*}[t]
\includegraphics[width=18cm]{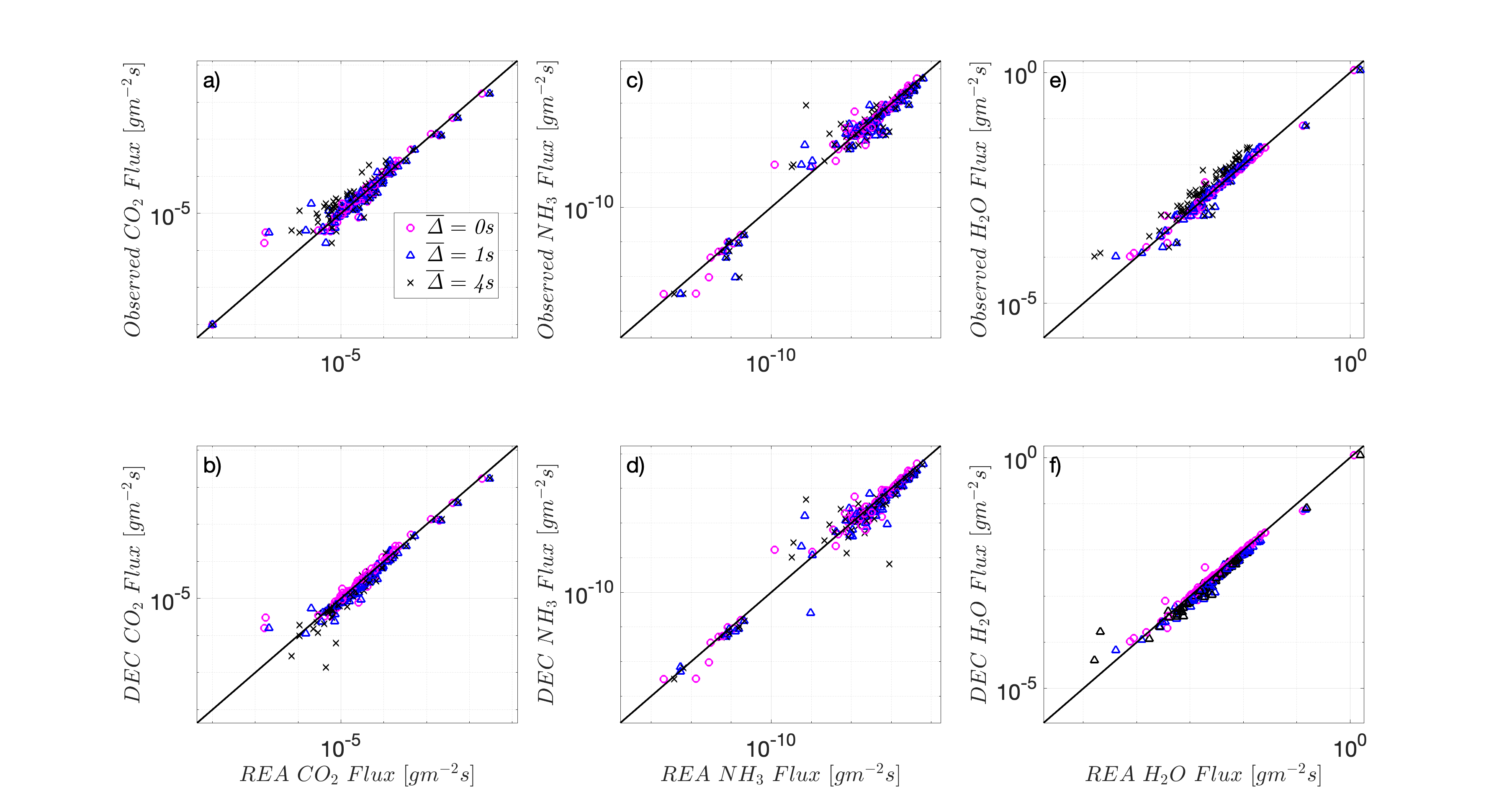}
\caption{ Similar to Fig. \ref{fig:5} but under unstable conditions.}
\label{fig:8}
\end{figure*}

Figs. \ref{fig:7} and \ref{fig:8} show the same plots as Figs. \ref{fig:4} and \ref{fig:5} respectively, but here under unstable conditions. As seen here, REA is insensitive to filtering operations.  

\noappendix       




\appendixfigures  

\appendixtables   


\authorcontribution{M.A. and E.B.Z. designed the framework of this study; M.A., E.B.Z, N.D., E.Z., G.G.K. developed the analysis methodologies; M.A. performed the formal analysis and visualization, V.I.S and M.A.Z. collected and curated the data from Wendell, working will April Leytem from the U.S. Department of Agriculture and John Walker and Ryan Fulgham from the U.S. Environmental Protection Agency. J.D.F. collected and curated the data from Utqiagvik; working with collaborating Environment Canada scientist Ralf Staebler. M.A. and E.B.Z. developed the first draft of the paper; all of the authors assisted in interpreting the results and contributed to writing the paper.} 

\competinginterests{The authors declare no competing interest.} 


\begin{acknowledgements}
M.A. and E.B.Z. are supported by the Cooperative Institute for Modeling the Earth System at Princeton University under Award NA18OAR4320123 from the National Oceanic and Atmospheric Administration, and by the US National Science Foundation under award number AGS 2128345.

V.I.S. is supported by the National Defense Science and Engineering Graduate Fellowship from the U.S. Department of Defense and Army Research Office.

M.A.Z. acknowledges the support from the U.S. Environmental Protection Agency (\#68HERH21D0006).

G.K. acknowledges the support from the U.S. National Science Foundation (NSF-AGS-2028633) and the Department of Energy (DE-SC0022072).

J.D.F. acknowledges the support provided by the National Science Foundation to complete the PHOXMELT field studies (Grant PLR- 1417914) to collect the data.

\end{acknowledgements}






\bibliographystyle{copernicus}
\bibliography{references}

\end{document}